\documentclass[lettersize,journal]{IEEEtran}
\usepackage{amsmath,amsfonts}
\usepackage{algorithm}
\usepackage{algpseudocode}
\usepackage{array}
\usepackage[caption=false,font=normalsize,labelfont=sf,textfont=sf]{subfig}
\usepackage{textcomp}
\usepackage{stfloats}
\usepackage{url}
\usepackage{verbatim}
\usepackage{graphicx}
\usepackage{cite}
\usepackage[font=small]{caption}
\usepackage{xcolor}
\usepackage{hyperref}
\usepackage{dsfont}
\usepackage{color,soul}
\usepackage[normalem]{ulem}
\begin{document}

\title{Data-Driven Stochastic Distribution System Hardening\\ Based on Bayesian Online Learning}

\author{Wenlong~Shi,~\IEEEmembership{Member,~IEEE},~Hongyi~Li,~\IEEEmembership{Member,~IEEE},~and~Zhaoyu~Wang,~\IEEEmembership{Senior  Member,~IEEE}
\thanks{This work was partially supported by the Power System Engineering and Research Center under Grant PSERC S-110, the U.S. Department of Energy Office of Energy Efficiency and Renewable Energy under Grant DE-EE0011234, and the National Science Foundation under Grant ECCS 2042314.}        
\thanks{The authors are with the Department of Electrical and Computer Engineering,
		Iowa State University, Ames, IA 50011 USA (e-mail:
		wshi5@iastate.edu; hongyili@iastate.edu; wzy@iastate.edu).}% <-this % stops a space
\thanks{(Corresponding author: Zhaoyu Wang.)}
}

% The paper headers
%\markboth{Journal of \LaTeX\ Class Files,~Vol.~14, No.~8, August~2021}%
%{Shell \MakeLowercase{\textit{et al.}}: A Sample Article Using IEEEtran.cls for IEEE Journals}

%\IEEEpubid{0000--0000/00\$00.00~\copyright~2021 IEEE}
% Remember, if you use this you must call \IEEEpubidadjcol in the second
% column for its text to clear the IEEEpubid mark.

\setlength{\abovedisplayskip}{3.5pt}
\setlength{\belowdisplayskip}{3.5pt}

\maketitle

\begin{abstract}
Extreme weather frequently cause widespread outages in distribution systems (DSs), demonstrating the  importance of hardening strategies for resilience enhancement. However, the well-utilization of real-world outage data with associated weather conditions to make informed hardening decisions in DSs is still an open issue. To bridge this research gap, this paper proposes a data-driven stochastic distribution line (DL) hardening strategy. First, a deep neural network (DNN) regression model is developed to predict the probabilistic evolution of outage scenarios under various hardening decisions. Based on the DNN predictions, the problem is formulated as a decision-dependent distributionally robust optimization (DRO) model, accounting for uncertainties in outage scenario distributions using a data-driven ambiguity set. To address decision-dependent uncertainty, a Bayesian online learning algorithm is proposed. This algorithm decomposes the original problem into inner and outer problems. Then, it iteratively refines hardening decisions by sequentially incorporating outage data and  dynamically updating decision-specific ambiguity sets by using Bayes’ theorem and Bayesian Inference. Also, the convergence of the algorithm is proven through dynamic regret analysis. Finally, case studies are implemented on a real-world DS in Redfield, Iowa, USA. A dataset spanning 24 years (2001–2024) is constructed based on the utility outage records. The simulation results validates the effectiveness of the proposed strategy.

\end{abstract}


\section*{Nomenclature}

\noindent\textit{A. Parameters}

\begin{description}[\IEEEsetlabelwidth{$(\cdot)_r + \textbf{i}(\cdot)_i$}]
	\item[$\boldsymbol{1}$]  Column vector with three  all $1$ entries.
    \item[$B^{\text{max}}$] Maximum available hardening budget.
    \item[$c_z^{pl},c_z^{ud}$] Hardening costs of  line segment $z$ in terms of pole upgrading and line undergrounding.
    \item[$c_z^{pd}$] Hardening cost of pad-mounted transformer.
    \item[$\boldsymbol{D}_i^p,\boldsymbol{D}_i^q$] Maximum Active and reactive load demand.
    \item[$\tilde{\boldsymbol{I}}_{ij}^\text{max}$] Maximum current capacity of line $(i,j)$.
    \item[$M$] A sufficiently large number fo big-$M$.
    \item[$\boldsymbol{P}_k^\text{max},\boldsymbol{Q}_k^\text{max}$] Rated active and reactive capacity of source $k$.
    \item[$\boldsymbol{r}_{ij},\boldsymbol{x}_{ij}$] Resistance and reactance matrices of line $(i,j)$.
    \item[$\tilde{\boldsymbol{V}}^{\text{min}},\tilde{\boldsymbol{V}}^{\text{max}}$] Maximum and minimum nodal voltage limits.\item[$\boldsymbol{z}_{ij}$] Impedance matrix of line $(i,j)$.
    %\item[$\mathcal{R}_{ij}$] Inrush current limits on switch $(i,j)$.
    \item[$\omega_i$] Load  weight at node $i$ for priority consideration.
    \item[$\alpha_z$] Binary indicator, 0 if faults occur on line segment $z$, otherwise 1. 
    \item[$\Delta t_s$] Outage duration under scenario $s$.
    
\end{description}

\noindent\textit{B. Variables}

\begin{description}[\IEEEsetlabelwidth{$(\cdot)_r + \textbf{i}(\cdot)_i$}]
    \item[$F_{ij}$] Fictitious commodity flow on line  $(i,j)$.
    \item[$h_z^{\text{ud}},h_z^{\text{pl}}$] Binary variables, 1 if undergrounding lines and upgrading poles are chosen for line segment $z$.
    \item[$\tilde{\boldsymbol{I}}_{ij}^{k}$] 3-phase squared current magnitude on line $(i,j)$ with respect to source $k$.
	\item[$\boldsymbol{p}_i,\boldsymbol{q}_i$] Active and reactive load demand at node $i$.
	\item[$\dot{\boldsymbol{ p}}_{ij}^{k}, \dot{\boldsymbol{ q}}_{ij}^{k}$] Slack variables for active and reactive power balance when switch $(i,j)$ is open.
	\item[$\boldsymbol{P}_{ij}^{k}, \boldsymbol{Q}_{ij}^{k}$] 3-phase active and reactive power flow on line $(i,j)$ with respect to source $k$.
    \item[$u_i^k$] Binary variable, equals 1 if node $i$ is activated by source  node $k$, otherwise 0.
    \item[$\dot{\boldsymbol{ v}}_{j}^{k}$] Slack variable for voltage at node $j$ when switch $(i,j)$ is disconnected.
    \item[$\tilde{\boldsymbol{V}}_{i}^{k}$] 3-phase squares of voltage magnitude of node
$i$ with respect to node $k$.
    \item[$\beta_i$] Binary vairable, equals 0 if the load at node $i$ is not served, otherwise 1. 
    \item[$\gamma_{ij}$] Binary variable, 0 if switch $(i,j)$ is open.
    \item[$\Gamma_{ij}$] Binary, 0 if line $(i,j)$ in the ficticious graph is disconnected to maintain radiality.
    \item[$\boldsymbol{\sigma}_{i,s}$] Load demand not being served at node $i$.

\end{description}

\noindent\textit{C. Indices and Sets}

\begin{description}[\IEEEsetlabelwidth{$(\cdot)_r + \textbf{i}(\cdot)_i$}]
    \item[$i,j$] Indices for nodes.
    \item[$(i,j)$] Indices for lines, switches, fuses, reclosers between nodes $i$ and $j$.
    \color{black}
    \item[$m$] Index of training dataset.
    \item[$k\in\mathcal{K}$] Index of source nodes in the network.
    \item[$s\in \mathcal{S}$] Index of outage scenarios.
    \item[$t$] Iteration index of the online algorithm.
    \color{black}
    \item[$z\in\mathcal{Z}$] Index of components for hardening.
    \item[$\mathcal{B}, \mathcal{B}_z$] Sets of switches and boundaries of  segment $z$.
    \item[$\mathcal{C}_i$] Set of child nodes of node $i$.
    \item[$\mathcal{U},\mathcal{P}$] Sets of geographic restrictions on undergrounding and pole upgrading.
    \item[$\mathcal{Z}_s,\mathcal{Z}_t$] Sets of components considering line segments and distribution transformers, respectively.
    \item[$\phi$] Index of phase in three-phase systems.

\end{description}

\section{Introduction}

\IEEEPARstart{E}{xtreme} weather events, such as hurricanes and severe storms, impose significant challenges to the resilience of power distribution systems (DSs). These natural disasters often result in widespread outages, disrupting electricity supply for extended durations \cite{shi2022enhancing}. For example, Hurricane Ian in 2022 caused 5-hours electricity interruption, which affected over 2.7 million customers in Florida, USA \cite{Ian}. And, the 2021 Texas winter storm resulted in 52,000 MW of power capacity being offline \cite{white2021method}. To mitigate these negative impacts, DS hardening measures such as undergrounding lines and upgrading poles are necessary. These measures are very  commonly adopted by utilities, which are proven effective in resilience enhancement.

However, obtaining which poles to upgrade and which line segments to underground is challenging. Hardening strategies relying on empirical analysis obtain hardening decisions based on iterative simulation and assessment \cite{ahmad2023towards,ahmad2024quantifying}. Despite the approaches are intuitive, their optimality cannot be guaranteed. In other words, theoretical analysis is necessary to deal with outage complexities. In this respect, optimization provides an effective way. In particular,  robust optimization (RO), stochastic programming (SP), and distributionally robust optimization (DRO) are useful in tackling uncertainties. For example, RO employs uncertainty sets to represent uncertain parameters. It is developed in a min-max form to obtain hardening decisions against the worst-case scenario   \cite{wang2017robust,li2023robust,he2018robust,lin2018tri,zhang2020multi,wang2022two,gan2021tri}. As RO contains only deterministic variables,
their performance can degrade when addressing the stochastic
nature of extreme events. In addition, SP considers stochasticity via scenario sets with probability distributions. It optimizes the expected cost value throughout all outage scenarios   \cite{ma2016resilience,hou2023resilience,chen2023two,tan2018distribution}. DRO constructs an ambiguity set of probability distributions to address the randomness. It considers the worst-case expected hardening performance   \cite{li2023distributionally}. However, both SP and DRO requires probability distributions, which should be accurate to avoid under- or over-estimation.

Modern power systems have transformed into more intelligent networks, driven by advanced sensing and communication technologies \cite{wang2014coordinated}. These advancements facilitate the collection of network conditions, making data-driven hardening strategies possible. For example, in \cite{bagheri2018resilient,bagheri2019distributionally,rahim2024wasserstein}, the statistics are extracted from outage records to establish data-driven ambiguity sets. Accordingly, a modified DRO is  developed such that hardening decisions can be obtained without relying on explicit models. Nonetheless, these research are predominantly investigated for transmission systems. They do not adapt to the unique features of outages and restoration at the distribution level.

In practice, DS operators monitor the network condition routinely via  Supervisory Control and Data Acquisition (SCADA) systems. The outage related information becomes available at the same time. For example, for isolating downstream faults, the clearing devices, such as fuses and reclosers installed along main feeders or at distribution transformers, are recorded. The event time stamp, restoration times, and affected customers are also documented  \cite{zhang2018big,nichelle2021extracting}. This dataset have the potential to support hardening decision-making. However, how to achieve this remains underexplored.

According to the aforementioned issue, this paper proposes a data-driven stochastic DL hardening strategy. Specifically, we employ DRO formulation to address the stochasticity, and we develop an online learning process to solve the data-driven challenge. The main contributions are as follows:

\begin{enumerate}
    \item A unique feature of  real-world outage data is that it is collected under fixed network conditions. To address this challenge, a deep neural network (DNN) regression model is proposed in this paper. By learning the nonlinear relationships between hardening measures and their impacts on outage outcomes, the model effectively captures the complex evolution of outage scenarios.

    \item For utilizing outage data from SCADA systems and weather information from  stations, and enable stochastic analysis, a data-driven stochastic DL hardening problem is proposed. The problem is formulated as a decision-dependent DRO model with a data-driven ambiguity set. Note that the problem can not be solved using traditional algorithms due to the decision-dependent uncertainty.

    \item To address the decision-dependent issue and solve the problem efficiently, a Bayesian online learning algorithm is developed. This algorithm first decomposes the original problem into inner and outer problems. Then, it refines hardening decisions in an iterative manner by sequentially learning from historical data and dynamically updating the ambiguity set based on Bayes' Theorem and Bayesian Inference. The convergence is also proven based on dynamic regret analysis.
    
\end{enumerate}

The remainder of this paper is structured as follows. Section II presents the system model, including the network model and the DNN regression model for outage prediction. Section III formulates the data-driven stochastic DL hardening problem as a decision-dependent DRO model. In Section IV, a Bayesian online learning algorithm is developed. Section V evaluates the proposed strategy through numerical experiments on a real-world distribution network in Redfield, Iowa, USA. Finally, Section VI concludes the paper and outlines future directions.

\begin{figure}[t]
    \centering	\includegraphics[width=3.4in]{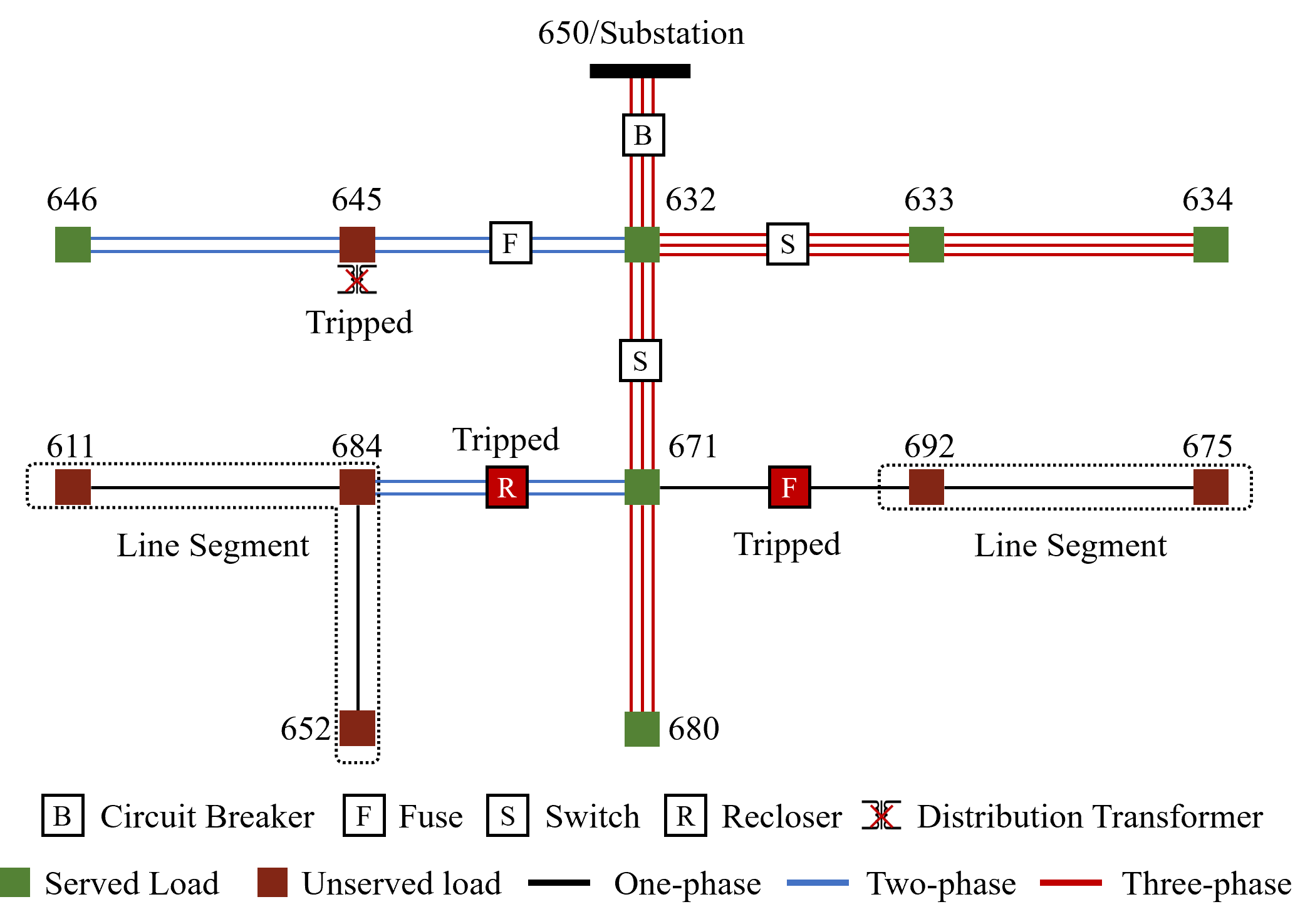}	
    \caption{An illustration of the three-phase unbalanced DS  model considering real-world outage data and protection devices.}
    \label{SystemModel}\vspace{-5pt}
\end{figure}
\section{System Model}

In this section, the distribution network model considering real-world outage data is presented. A DNN regression model is proposed to predict the probability distribution of outage scenarios under varying hardening decisions.

\vspace{-5pt}

\subsection{Distribution Network Model}

In this paper, the distribution network model is developed to facilitate the utilization of real-world outage data. Specifically, we consider three-phase unbalanced systems. The distribution network is modeled as a graph $\mathcal{G} = (\mathcal{N}, \mathcal{L})$, where $\mathcal{N}$ represents the set of nodes and $\mathcal{L}$ represents the set of lines. A line between two nodes can represent a DL, a switch or a protection device. In addition, the protection devices are assumed to install along main feeders, including circuit breakers, reclosers, and fuses to isolate faults on main feeders. We also have fuses installed at distribution transformers to clear faults on laterals. Then, the portion of lines between two or more protection devices, is defined as a line segment. To enhance system resilience against extreme weather events, hardening measures including upgrading poles and undergrounding lines are applied to line segments. The deployment of pad-mounted transformers is employed to protect distribution transformers. Beyond protection devices, switches are also installed along main feeders for  topology reconfiguration \cite{wang2015self}.

An example based on IEEE 13-node test feeder is shown in Fig. \ref{SystemModel}. Specifically, a recloser and a fuse are tripped to isolate the faults downstream of node 684 and 692, respectively. Also, the distribution transformer at node 645 is tripped to isolate the fault on laterals. Furthermore, the faults are represented as a set of outage scenarios. The outage scenarios are classified into two types. First, when a fault happens on the main feeder, it can be easily located to one specific line segment between  protection devices. However, the exact fault location within this line segment does not influence the automated operation of the protection devices \cite{dos2012simultaneous}. Consequently, each line segment is corresponding to an outage scenario. Second, when a fault occurs on the lateral, the fuse at the distribution transformer will trip to isolate the fault. This defines another type of outage scenarios that isolate the downstream of the transformer.

\vspace{-5pt}
\subsection{DNN Regression Model}

For DL hardening problem, a significant challenge lies in the fact that outage data is predominantly collected under existing and fixed network conditions. In practice, it is nearly impossible to obtain outage data corresponding to the multitude of potential hardening solutions, as implementing and testing these solutions across diverse weather conditions require impractical levels of time, resources, and infrastructure. To overcome this limitation, a DNN regression model is proposed. The objective is to predict how a real-world outage scenario, that is observed under fixed network conditions, will evolve when subjected to hypothetical hardening measures. 

We first construct the training dataset. Specifically, considering an outage scenario contains a set of $\mathcal{Z}'$ fault components including line segments and distribution transformers. \textcolor{black}{We use $\mathcal{I}_z\in[0,1]$ to represent the improvement in the failure probability of component $z$ if hardening measures are taken \cite{darestani2019multi}. The improvement is integrated into the dataset by assigning the scenario with a label indicating an occurrence of $\prod_{z\in\mathcal{Z}'}(1 - \mathcal{I}_z)$. The remaining probability $1-\prod_{z\in\mathcal{Z}'}(1 - \mathcal{I}_z)$ is evenly distributed across the other scenarios with fault components $z'\in\mathcal{Z}'\backslash z$ for normalization. Such an adjustment assumes that the effects of hardening different components are independent. It is practical, as physically reinforcing one component does not influence the failure probability of others in most distribution systems.} The training dataset can be 
represented as $\mathcal{T}=\{(\boldsymbol{h}_m,\boldsymbol{c}_m,\boldsymbol{o}_m,\boldsymbol{o}'_m)\}_{m=1}^M$, where $m$ is the instance index. Specifically, $\boldsymbol{h}_m = [h_{m,1}, h_{m,2}, \dots, h_{m,|\mathcal{Z}|}] \in \{0, 1\}^{|\mathcal{Z}|}$ represents a binary vector with $|\mathcal{Z}|$ components, each corresponding to a hardening decision $h_{m,z}$, where $h_{k,z}=1$ when component $z$ is hardened. $\boldsymbol{c}_m = [c_{m,1}, c_{m,2}, \dots, c_{m,|\mathcal{M}|}] \in \mathbb{R}^{|\mathcal{M}|}$ contains $|\mathcal{M}|$  weather parameters, such as wind speed, humidity and temperature. Also, $\boldsymbol{o}_m = [o_{m,1}, o_{m,2}, \dots, o_{m,|\mathcal{S}|}] \in \{0, 1\}^{|\mathcal{S}|}$ is a binary vector for outage scenarios, where $o_{m,s}=1$ if  scenario $s$ occurs. And, $\boldsymbol{o}'_m = [o'_{m,1}, o'_{m,2}, \dots, o'_{m,|\mathcal{S}|}] \in [0, 1]^{|S|}$ denotes the outage occurrence, serving as labels for learning and making predictions. Furthermore, the DNN model consists of an input layer, hidden layers, and an output layer. The input layer vector is $\{\boldsymbol{h},\boldsymbol{c},\boldsymbol{o}\}$ including $|\mathcal{Z}|+|\mathcal{M}|+|\mathcal{S}|$ neurons. The output layer contains $|S|$ neurons each corresponding to the probability of an outage scenario. In addition, to ensure the output of the DNN sums up to 1, the softmax function is applied as the activation function of the output layer. Note that the trained DNN serves as a regression model for scenario translation. It can be described as a function $f_w: (\boldsymbol{h}, \boldsymbol{c},\boldsymbol{o}) \to \boldsymbol{o}'$, where $w$ represents the learnable parameters. It transforms data collected under fixed network conditions into predictions to reflect the effects of hypothetical  hardening decisions. \textcolor{black}{Note that, in this paper, the DNN is trained on outage records associated with wind-related weather covariates. The network architecture is inherently flexible and not restricted to specific input types. Additional ambient variables, such as temperature, precipitation, ice accretion, or lightning frequency, can be included into the input feature vector. As a result, the DNN can be adapted to other geographic regions where different climatic factors are more prominent, provided that corresponding historical data are available for training.}

\vspace{-5pt}
\section{Data-Driven Stochastic DL Hardening Problem Formulation}
In this paper, we aim to bridge the research gap by leveraging long-term historical outage data to inform and optimize the stochastic DL hardening decision against extreme weather. To this end, the problem is formulated as a decision-dependent 
DRO model with a data-driven ambiguity set. In the rest of this section, the problem constraints are presented, and the problem objective is discussed in details.

\vspace{-5pt}
\subsection{Problem Constraints}

In this section, the problem constraints are presented in detail. Note that except for the investment constraints, all  constraints are scenario-specific. However, for notation simplicity, outage scenario $s$ is not explicitly indicated in each equation. 

\subsubsection{Operational Constraints} To ensure an efficient  operation for  restoration, the following constraints are applied:
\begin{equation}\label{O1}
    \gamma_{ij}  \leq \alpha_{z}, \forall (i,j)\in\mathcal{B}_z, z\in\mathcal{Z}_s,
\end{equation}
\begin{equation}\label{O2}
    \beta_i  \leq \alpha_{z},  i=z, \forall z\in\mathcal{Z}_t,
\end{equation}
\begin{equation}\label{O3}
    0 \leq \boldsymbol{p}_i \leq \beta_i\boldsymbol{D}_i^p, 0 \leq \boldsymbol{q}_i \leq \beta_i\boldsymbol{D}_i^q,  \forall i\in\mathcal{N},
\end{equation}
\begin{equation}\label{O4}
    \boldsymbol{\sigma}_i = \boldsymbol{D}_i^p - \boldsymbol{p}_i,  \forall i\in\mathcal{N},
\end{equation}
\begin{equation}\label{O5}
    u_k^k = 1, \forall k\in\mathcal{K},
\end{equation}
\begin{equation}\label{O6}
\sum\nolimits_{k\in\mathcal{K}}u_i^{k}\leq 1, \forall i\in\mathcal{N},
\end{equation}
\begin{equation}\label{O7}
    u_j^k \leq u_i^k, \forall j\in\mathcal{C}_i,
\end{equation}
\begin{equation}\label{O8}
    u_i^k + u_j^k \leq 1+\gamma_{ij}, \forall (i,j)\in\mathcal{B},k\in\mathcal{K},
\end{equation}
\begin{equation}\label{O9}
    |u_i^k - u_j^k| \leq 1-\gamma_{ij}, \forall (i,j)\in\mathcal{B},k\in\mathcal{K}.
\end{equation}

Constraint (\ref{O1}) models fault isolation, implying that if faults occur on line segment $z$, the switches on the boundaries must be open. Constraint (\ref{O2}) means that the load can not be served if the distribution transformer is tripped due to faults on laterals. Constraints (\ref{O3})-(\ref{O4}) are for unserved loads \cite{wang2014time}. Constraints (\ref{O5})-(\ref{O6}) imply  one node can only be activated by one source node for voltage and frequency regulation. Constraint (\ref{O7}) means that a child node can be activated only when its parent node is activated. Constraints (\ref{O8})-(\ref{O9}) ensure two ends of a switch $(i,j)$ are disconnected if the switch is open. In addition, if line $(i,j)$ is hard connected, meaning it is non-switchable, the activation status of nodes $i$ and $j$ must remain consistent.

\subsubsection{Power Flow Constraints}

In this paper, the branch flow model is adopted as constraints to provide a set of closed-form power flow equations \cite{farivar2013branch}. The three-phase active and reactive power balance equations can be given by
\begin{equation}\label{PF1}
    \textstyle\sum_{j'\in\mathcal{C}_j}\boldsymbol{P}_{jj'}^{k}=\boldsymbol{P}_{ij}^{k}-\boldsymbol{p}_{j}-\boldsymbol{r}_{ij}{\tilde{\boldsymbol{I}}_{ij}^{k}} + \dot{\boldsymbol{ p}}_{ij}^k, \forall (i,j)\in\mathcal{L},
\end{equation}
\begin{equation}\label{PF2}
    \textstyle\sum_{j'\in\mathcal{C}_j}\boldsymbol{Q}_{jj'}^{k}=\boldsymbol{Q}_{ij}^{k}-\boldsymbol{q}_{j}-\boldsymbol{x}_{ij}{\tilde{\boldsymbol{I}}_{ij}^{k}}+ \dot{\boldsymbol{ q}}_{ij}^k, \forall (i,j)\in\mathcal{L},
\end{equation}
where $\tilde{\boldsymbol{I}}_{ij}^{k}$ is the squared magnitude of the current from node $i$ to $j$ with respect to the source $k$. In addition, the power flow constraints after angle and conic relaxations can be given by
\begin{eqnarray}\label{PF3}
	\tilde{\boldsymbol{I}}_{ij}^{k}\odot{\tilde{\boldsymbol{V}}_{i}^{k}}\geq (\boldsymbol{P}_{ij}^{k})^{\odot 2}+(\boldsymbol{Q}_{ij}^{k})^{\odot 2}, \forall (i,j)\in\mathcal{L},
\end{eqnarray}
\begin{equation}\label{PF4}
	{\tilde{\boldsymbol{V}}_{i}^{k}}-{\tilde{\boldsymbol{V}}_{j}^{k}}=2(\hat{\boldsymbol{r}}_{ij}\boldsymbol{P}_{ij}^{k}\!+\!\hat{\boldsymbol{x}}_{ij}\boldsymbol{Q}_{ij}^{k}) +|\hat{\boldsymbol{z}}_{ij}|^2{\tilde{\boldsymbol{I}}_{ij}^{k}}+{\dot{\boldsymbol{ v}}_{j}^{k}},
\end{equation}
where $\hat{\boldsymbol{z}}_{ij}=\hat{\boldsymbol{x}}_{ij}+j\hat{\boldsymbol{r}}_{ij}$ is the equivalent impedance matrix \cite{nejad2019distributed}. Notice that constraint (\ref{PF3}) is a second-order cone constraint which is convex. The terms $\dot{(\cdot )}$ denote slack variables, such that constraint (\ref{PF1})-(\ref{PF4}) can be feasible if switch $(i,j)$ is opened. To achieve this, the following constraints are  applied:
\begin{eqnarray}\label{PF5}
	\boldsymbol{0}\leq{\dot{\boldsymbol{ p}}_{ij}^{k}},{\dot{\boldsymbol{ q}}_{ij}^{k}}\leq ({1}-u_j^k)\boldsymbol{M}, \forall (i,j)\in\mathcal{B},
\end{eqnarray}
\begin{eqnarray}\label{PF5}
	\boldsymbol{0}\leq{\dot{\boldsymbol{v}}_{j}^{k}}\leq ({1}-u_j^k)\boldsymbol{M}, \forall j\in\mathcal{C}_i.
\end{eqnarray}

Moreover, we have the following constraints for the nodes if they are not restored by the $k_{th}$ source node \cite{ma2022robust}:
\begin{equation}\label{PF6}
	\sum\nolimits_{k\in\mathcal{K}}u_i^{k}\tilde{\boldsymbol{V}}^\text{min}\leq\sum\nolimits_{k\in\mathcal{K}}\tilde{\boldsymbol{V}}_{i}^{k}\leq \sum\nolimits_{k\in\mathcal{K}}u_i^{k}\tilde{\boldsymbol{V}}^\text{max}, 
\end{equation}
\begin{equation}\label{PF7}
	\boldsymbol{0} \leq \boldsymbol{P}_{ij}^{k} \leq u_j^k\boldsymbol{P}_k^\text{max}, \forall j\in\mathcal{C}_i, k\in\mathcal{K},
\end{equation}
\begin{equation}\label{PF8}
	\boldsymbol{0} \leq \boldsymbol{Q}_{ij}^{k} \leq u_j^k\boldsymbol{Q}_k^\text{max}, \forall j\in\mathcal{C}_i, k\in\mathcal{K},
\end{equation}
\begin{equation}\label{PF9}
    0\leq\tilde{\boldsymbol{I}}_{ij}^{k}\leq u_j^k\tilde{\boldsymbol{I}}_{ij}^\text{max}, \forall j\in\mathcal{C}_i, k\in\mathcal{K}.
\end{equation}

Constraint (\ref{PF6}) means  that if node $i$ is not activated by any source $k$, its voltage should be $0$. This constraint also ensures that all the nodal voltages remain within the specified limits if the nodes are activated. Constraints (\ref{PF7})-(\ref{PF9}) enforce power flow and current on line $(i,j)$ to $0$, if node $j$ is not powered by source $k$. Note that Constraints (\ref{PF7})-(\ref{PF8}) also limit the output of source $k$ within rated capacity.

\subsubsection{Radial Network  Constraints} The operation of tie-lines can introduce potential loops. Hence, the following constraints based on single commodity flow model are essential to maintain a spanning-tree after topology reconfiguration \cite{ding2017new}. 
\begin{equation}\label{R1}
    \sum\nolimits_{i \in \pi_j} F_{ji} - \sum\nolimits_{i \in \pi_j} F_{ij} = -1,  \forall j \notin \mathcal{K},
\end{equation}
\begin{equation}\label{R2}
    \sum\nolimits_{i \in \pi_k} F_{ki}  \geq 1,  \forall k \in \mathcal{K},
\end{equation}
\begin{equation}\label{R3}
\sum\nolimits_{(i,j)\in\mathcal{B} }\beta_{ij} = |\mathcal{N}| - |\mathcal{K}|,
\end{equation}
\begin{equation}\label{R4}
    -\Gamma_{ij}M \leq F_{ij}\leq \Gamma_{ij}M, \forall (i,j)\in\mathcal{B},
\end{equation}
\begin{equation}\label{R5}
    \gamma_{ij} \leq \Gamma_{ij}, \forall (i,j)\in\mathcal{B}.
\end{equation}

Constraint (\ref{R1})-(\ref{R3}) ensures radial connectivity by requiring that exactly one unit of a fictitious commodity flows from the source node to every other node in the network. Constraint (\ref{R4}) enforces the fictitious flow on line $(i,j)$ to $0$ if it is disconnected. These constraints achieves radiality by duplicating the distribution network into a fictitious network with the same topology. Ensuring radiality in the fictitious network inherently guarantees radiality in the original network. Constraint (\ref{R5}) ensures that a switch can only be operated when the radiality condition is satisfied.

\subsubsection{Investment Constraints} To ensure that the hardening decisions in the DS are financially viable and in accordance with the available budget, the following constraints are applied:
\begin{equation}\label{BG1}
    \sum\nolimits_{z\in\mathcal{Z}_s} \left( c_z^{\text{pl}} h_z^{\text{pl}} + c_z^{\text{ud}} h_z^{\text{ud}} \right) + \sum\nolimits_{z\in\mathcal{Z}_t} {c_z^{\text{pd}} h_z^{\text{pd}}} \leq B^{\text{max}},
\end{equation}
\begin{equation}\label{BG2}
    h_z^{\text{pl}} + h_z^{\text{ud}} \leq h_z, z\in \mathcal{Z}_s,
\end{equation}
\begin{equation}\label{BG3}
    h_z^{\text{pl}}\leq 0, z\in \mathcal{P} \enspace\text{or} \enspace h_z^{\text{ud}} \leq 0, z\in \mathcal{U}.
\end{equation}

Constraint (\ref{BG1}) ensures the total investment must be within the budget. Constraint (\ref{BG2}) stipulates  that each line segment can be hardened by only one measure. Constraint (\ref{BG3}) includes geographic requirements based on local conditions, allowing only pole upgrading or line  undergrounding in certain areas.
\vspace{-10pt}
\subsection{Problem Objective}
The data-driven stochastic DL hardening problem is formulated as a decision-dependent DRO model. The objective is to minimize the worst-case expected unserved load over an ambiguity set of probability distributions:
\begin{equation}
    (P1)\quad \min_{\boldsymbol{h} \in \mathcal{H}} \max_{p \in \mathcal{P}} \mathbb{E}_{s \sim p}[f(\boldsymbol{h}, s)],
\end{equation}
where $\mathbb{E}_{s \sim p}[f(\boldsymbol{h}, s)]$
denotes the expected cost over all possible scenarios. For a finite outage scenario set, the expectation simplifies to $\mathbb{E}_{s \sim p} \left[ f(\boldsymbol{h}, s) \right] = \sum_i p_i f(\boldsymbol{h}, s_i)$ based on sample average approximation (SAA), where $f(\boldsymbol{h}, s)$ is given by
\begin{equation}
    f(\boldsymbol{h}, s) = f_1(\boldsymbol{h}, s)+f_2(\boldsymbol{h}, s),
\end{equation}
\begin{equation}
    f_1(\boldsymbol{h}, s) = \sum\nolimits_{i\in\mathcal{N}, \phi\in\{a,b,c\}}\omega_i\boldsymbol{\sigma}_{i,s}\Delta t_s,
\end{equation}
\begin{equation}
    f_2(\boldsymbol{h}, s) = \sum\nolimits_{(i,j)\in\mathcal{L}, \phi\in\{a,b,c\}}(\boldsymbol{r}_{ij}\sum\nolimits_{k\in\mathcal{K}}\tilde{\boldsymbol{I}}_{ij}^{k}),
\end{equation}
where $f_1(\boldsymbol{h}, s)$ is the total weighted unserved load function, with $\boldsymbol{\sigma}_{i,s}\Delta t_s$ denoting the unserved load at node $i$ under scenario $s$. Also, $f_2(\boldsymbol{h}, s)$ is the total power losses function. This term is added such that the exactness of convex relaxation of the second-order cone constraint (\ref{PF3}) is guaranteed \cite{nejad2019distributed}.

To model the uncertainty in the probability distribution of outage scenarios, an ambiguity set $\mathcal{P}$ is constructed around the empirical distribution $\hat{\boldsymbol{p}}$. This set uses the $\ell_2$-norm to define the proximity to the empirical distribution:
\begin{equation}
    \mathcal{P} = \{ \boldsymbol{p} \in \mathbb{R}^n : \| \boldsymbol{p} - \hat{\boldsymbol{p}} \|_{2} \leq d \},
\end{equation}
where $\boldsymbol{p}$ denotes all potential distributions of damage scenarios within a distance $d$ of the empirical distribution. The parameter 
$d$ captures the uncertainty, reflecting the unmodeled variability in the DNN regression model. Note that the empirical distribution $\hat{\boldsymbol{p}}$ can be extracted from historical data, while it varies with respect to  hardening decision $\boldsymbol{h}$, making the DRO model decision dependent. Hence, it cannot be directly solved using traditional approaches such as Benders' decomposition.

\textcolor{black}{The $\ell_2$-norm is selected for both statistical and computational reasons. Statistically, it can offer finite sample coverage guarantees for empirical distributions. Computationally, it enables closed-form gradient evaluation and  projection via a simple bisection method with complexity $\mathcal{O}(|\mathcal{S}|)$. Its smoothness and convexity also facilitate dynamic regret analysis. While the Wasserstein distance provides advantages in continuous or spatially structured spaces, its application requires solving large-scale linear programs with $\mathcal{O}(|\mathcal{S}|^2)$ constraints. However, our strategy is not restricted to the Euclidean metric. If future applications require more complicated representations of uncertainty, such as Wasserstein-based ambiguity sets, the framework can be extended by modifying the projection step, without requiring fundamental changes to the overall structure.}

\begin{figure*}[t]
    \centering	\includegraphics[width=6.7in]{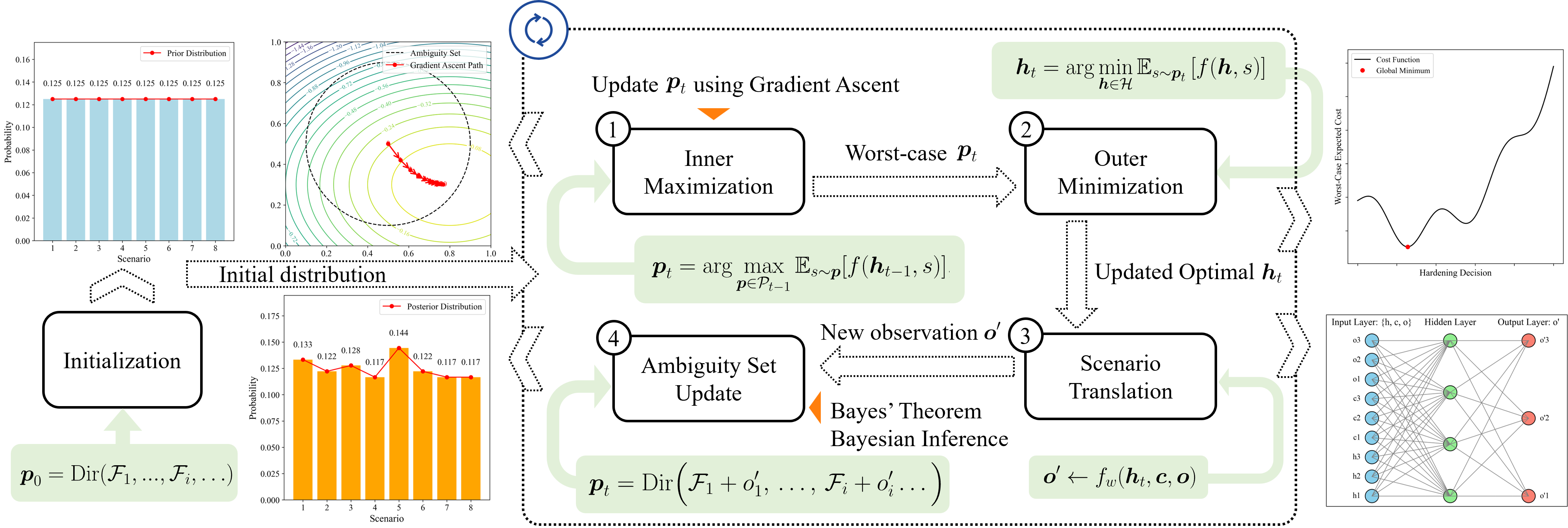}	
    \caption{An illustration of the proposed Bayesian online learning algorithm.}
    \label{Algorithm}\vspace{-5pt}
\end{figure*}

\vspace{-10pt}
\section{Data-Driven Stochastic DL Hardening Problem Solution}

To address the decision-dependent uncertainty and enhance computational efficiency, we propose a Bayesian online learning algorithm. As illustrated in Fig. \ref{Algorithm}, the algorithm iteratively refines hardening decisions by sequentially incorporating outage data. Specifically, at each iteration, the algorithm employs the proposed DNN regression model to translate an observed outage instance obtained  under fixed network conditions into a probability distribution that quantifies how a specific hardening decision alters outage likelihoods. The DNN prediction is then used to update the posterior distribution via Bayes' Theorem, yielding a refined empirical distribution that acts as the center of a decision-specific ambiguity set. After that, the original decision-dependent DRO problem is decomposed into an inner problem and an outer problem. The inner problem utilizes projected gradient ascent to identify the worst-case probability distribution. The outer problem updates the hardening decision by minimizing the worst-case expected unserved load. Then, the posterior distribution is employed as the global prior for the next iteration. Through the online learning, the underlying outage uncertainties are iteratively refined, and the hardening decision ultimately converges to robust solutions that capture real-world outage complexities. 

\textcolor{black}{Even though integrating multiple components, the proposed algorithm ensures robustness and stability. First, the DNN predicts the probabilistic distribution over outage scenarios, with prediction errors accounted for through the ambiguity set. Second, the DRO model, convex in the scenario probabilities, provides both the worst-case expected performance and its gradient to the online learning. It ensures that even a single projected gradient ascent step suffices for useful updates, without solving the inner loop to full optimality. Third, ambiguity sets are updated online via Dirichlet counts based on observed scenario occurrences, without requiring DNN retraining.}

\vspace{-5pt}
\subsection{Bayesian Online Learning Algorithm}

The pseudo code for the algorithm is presented in Algorithm \ref{A1}. Main steps of the algorithm are explained as follows.

\subsubsection{Initial Setup and Ambiguity Set Definition}
At the beginning of the learning process, initial hardening decisions are arbitrarily chosen to provide a starting point for the algorithm. An initial prior distribution is defined as:
\begin{equation}\label{LK}
    \boldsymbol{p}_0 = \text{Dir}(\mathcal{F}_1,...,  \mathcal{F}_i, \ldots,  \mathcal{F}_{|\mathcal{S}|}),
\end{equation}
where $\text{Dir}(\cdot)$ denotes the Dirichlet distribution, and $\mathcal{F}_i$ represents the initial cumulative fractional count for scenario $s_i$. Before any data is learned, $\boldsymbol{p}_0$ can be initialized as uniform distribution by having all $\mathcal{F}_i=1$. It can also be initialized based on expert knowledge. Note that $\boldsymbol{p}_0$ acts as the empirical distribution of the initial ambiguity set. And, as the algorithm progresses, the $\boldsymbol{\mathcal{F}}$ will be updated by adding fractional counts derived from the DNN’s output.

\subsubsection{Inner Maximization} Based on hardening decision at last step $\boldsymbol{h}_{t-1}$, the inner problem is solved to identify the worst-case probability distribution within the ambiguity set $\mathcal{P}_{t-1}$ that maximizes the expected cost, given by 
\begin{equation}
    (P2)\quad\boldsymbol{p}_t = \arg \max_{\boldsymbol{p} \in \mathcal{P}_{t-1}} \mathbb{E}_{s \sim \boldsymbol{p}}[f(\boldsymbol{h}_{t-1}, s)],
\end{equation}
where $\boldsymbol{p}_t$ represents the most adverse probability corresponding to the given decision $\boldsymbol{h}_{t-1}$. Since 
$\boldsymbol{p}\in\mathcal{P}_{t-1}$ is a probability distribution over a finite scenario set $\mathcal{S}$, the inner problem is linear in $\boldsymbol{p}$. \color{black}Consequently, the inner problem is differentiable w.r.t  
$\boldsymbol{p}$, making it suitable to the gradient ascent method:
\begin{equation}\label{GD}
\tilde{\boldsymbol{p}}_t = \boldsymbol{p}_{t-1} + \eta \nabla_{\boldsymbol{p}} \mathbb{E}_{s \sim \boldsymbol{p}_{t-1}} \left[ f(\boldsymbol{h}_{t-1}, s) \right],
\end{equation}
where $\eta$ is a small step size, and $\tilde{\boldsymbol{p}}_t$ is an approximation of $\boldsymbol{p}_t$. Equation (\ref{GD}) allows for adjustment of $\boldsymbol{p}$ in the direction of the gradient, which increases the expected cost for maximization. Note that $\tilde{\boldsymbol{p}}_t$ is an intermediate distribution, which may not yet satisfy the constraints of the ambiguity set, such as $\tilde{\boldsymbol{p}}_t(s_i) \geq 0$, $\sum_{i} \tilde{\boldsymbol{p}}_t(s_i)= 1$, and $\| \tilde{\boldsymbol{p}}_t - \boldsymbol{p}_{t-1 }\|_2 \leq d_{t-1}$. After gradient ascent, if $\tilde{\boldsymbol{p}}_t$ violates the constraints even after normalization, it is therefore projected back onto the ambiguity set $\mathcal{P}_{t-1}$ to ensure feasibility. This projection can be achieved by solving a convex quadratic programming problem:
\begin{equation}\label{GD1}
 \boldsymbol{p}_t = \arg \min_{\boldsymbol{p} \in \mathcal{P}_{t-1}} \frac{1}{2} \| \boldsymbol{p} -  \tilde{\boldsymbol{p}}_t \|_2^2,  
\end{equation}
where $\| \boldsymbol{p} -  \tilde{\boldsymbol{p}}_t \|_2^2$ is the regularization term utilized to prevent large jumps from $\boldsymbol{p}_{t-1}$. By combining equations (\ref{GD})-(\ref{GD1}), the probability $\boldsymbol{p}_t$ for the next iteration can be optimized by a compact-form optimization problem which simultaneously considers the gradient ascent and regularization, given by
\begin{equation}
   \min_{\boldsymbol{p} \in \mathcal{P}_{t-1}} \left\langle \eta \nabla_{\boldsymbol{p}} \mathbb{E}_{s \sim \boldsymbol{p}_{t-1}} \left[ f(\boldsymbol{h}_{t-1}, s) \right], \boldsymbol{p} \right\rangle + \frac{1}{2} \| \boldsymbol{p} - \boldsymbol{p}_{t-1} \|_2^2
\end{equation}
\begin{equation}\nonumber
  s.t.\enspace \boldsymbol{p} \geq \boldsymbol{0}, \left\langle \boldsymbol{p}, \boldsymbol{1} \right\rangle = 1, \| \boldsymbol{p} - \boldsymbol{p}_{t-1 }\|_2 \leq d_{t-1}
\end{equation}
where the first inner product term $\left\langle\cdot,\cdot\right\rangle$ represents how much the updated 
$\boldsymbol{p}_t$ aligns with the gradient direction\color{black}. \textcolor{black}{Notice that convergence of the inner problem is guaranteed as long as $\boldsymbol{p}$ is updated in a valid ascent direction at each iteration. Solving the inner maximization problem to full optimality is not required. Even though performing multiple projected gradient steps per iteration may yield a tighter approximation of the worst-case distribution, it is unnecessary in practice considering the trade-off between convergence and computational efficiency.}

These three equations collectively describe the same projected gradient update procedure, but from different and complementary perspectives. Equation (35) describes an unconstrained gradient ascent step, which updates the scenario probability vector in the direction of the gradient without enforcing any feasibility constraints. As a result, the intermediate update $\tilde{\boldsymbol{p}}_t$ may fall outside the feasible region. Equation (36) then introduces a projection step, which maps $\tilde{\boldsymbol{p}}_t$ back onto the admissible set defined by the $\ell_2$-norm trust region around the previous iteration. This projection not only ensures feasibility but also serves as a regularization mechanism that stabilizes the update and limits the step size. Finally, equation (37) presents the whole projected gradient update into a single constrained optimization problem, explicitly incorporating non-negativity, normalization, and proximity constraints. While equation (37) alone is sufficient from an algorithmic standpoint, we remain equations (35)-(36) and their respective explanations to make the underlying “gradient-then-project” logic transparent. This helps readers better understand how the direction of update is first determined and then regularized through projection. 

\subsubsection{Outer Minimization}
Using the worst-case distribution $\boldsymbol{p}_t$, the outer problem is further solved to update the hardening decision $\boldsymbol{h}_t$. The goal is to minimize the expected cost under the identified worst-case scenario, given by
\begin{equation}
    (P3)\quad\boldsymbol{h}_t = \arg \min\nolimits_{\boldsymbol{h} \in \mathcal{H}} \mathbb{E}_{s \sim \boldsymbol{p}_t} \left[ f(\boldsymbol{h}, s) \right].
\end{equation}

\subsubsection{Scenario Translation} 
Once the DL hardening $\boldsymbol{h}_t$ decision at step $t$ is obtained, it is input into the DNN regression model along with randomly selected outage data instance:
\begin{equation}
    \boldsymbol{o}' \leftarrow f_w (\boldsymbol{h}_t, \boldsymbol{c},\boldsymbol{o}).
\end{equation}
where $\boldsymbol{o}'$ denotes the DNN's predicted probability distribution over outage scenarios at the current iteration. This process directly addresses decision-dependent uncertainty, simulating the transformation of outage scenarios from data collected under fixed network conditions to those expected under hypothetical hardening decision $\boldsymbol{h}_t$.

\subsubsection{Ambiguity Set Update} 
At each iteration, the probability distribution over all scenarios $s$ output by the DNN model is used to update the likelihood of outage scenarios, given by
\begin{equation}\label{LK_soft}
    p(D_t \mid \boldsymbol{p}_{t-1}, \boldsymbol{h}_t) = \Big[o'_1,\, o'_2,\, \dots,\, o'_{|\mathcal{S}|}\Big],
\end{equation}
where $o'_s$ represents the fractional counts of scenarios, serving as the newest observation of the underlying outage uncertainty. Accordingly, by using Bayes' Theorem, the posterior distribution is updated as
\begin{equation}\label{BSTH_soft}
     \boldsymbol{p}_{t}(D_t, \boldsymbol{h}_t) \propto p(D_t \mid \boldsymbol{p}_{t-1}, \boldsymbol{h}_t) \boldsymbol{p}_{t-1}.
\end{equation}

Because the Dirichlet distribution is the conjugate prior for the multinomial distribution, the posterior distribution remains the Dirichlet form, with its parameters updated by adding the new fractional counts \(o'_i\) to the existing counts, given by
\begin{equation}\label{Bayes}
     \boldsymbol{p}_{t}(D_t, \boldsymbol{h}_t) = \text{Dir}\Big(\mathcal{F}_1 + o'_1,\, \mathcal{F}_2 + o'_2,\, \ldots,\, \mathcal{F}_{|\mathcal{S}|} + o'_{|\mathcal{S}|}\Big),
\end{equation}
where $\mathcal{F}_i$ denotes the fractional counts for $s_i$ from previous iterations. \textcolor{black}{The Dirichlet prior is adopted, as it offers a natural and analytically tractable approach for modeling uncertainty over discrete scenario probabilities. Its conjugacy with the categorical likelihood facilitates efficient Bayesian updating as new scenario observations are acquired.} Equation \ref{Bayes} imply that the new observation $\boldsymbol{o'}$ is governed by the prior belief $\boldsymbol{\mathcal{F}}$, which reflects the sequential learning process of the proposed algorithm. Then, by applying the expectation for the Dirichlet distribution, we have
\begin{equation}
    \mathbb{E}[p(s_i)] = \frac{\mathcal{F}_i + o'_i}{\sum_{j=1}^{|\mathcal{S}|} \Big(\mathcal{F}_j + o'_j\Big)}.
\end{equation}

Note that $\mathbb{E}[p(s_i)]$ provides the current best probability estimation for each scenario $s_i$. Therefore, based on the posterior mean, the decision-specific ambiguity set is constructed:
\begin{equation}
    \mathcal{P}_{t+1}(\boldsymbol{h}_t)= \{ \boldsymbol{p} \in \mathbb{R}^n : \| \boldsymbol{p} - \overline{\boldsymbol{p}}_t]\|_{2} \leq d_{t+1} \}.
\end{equation}
where $\overline{\boldsymbol{p}}_t=\big[\mathbb{E}[p(s_1)],\,\mathbb{E}[p(s_2)]\, \ldots,\,\mathbb{E}[p(s_{|\mathcal{S}|})]\big]$. Also, \(d_{t+1}\) at the current step reflects our confidence level in the updated posterior in representing the empirical distribution. \textcolor{black}{Note that the term “decision-specific ambiguity set” refers to the fact that the posterior update is conditioned on the hardening decision implemented. Specifically, at iteration $t$, a hardening vector $\boldsymbol{h}_t$ is selected, and the DNN is utilized to infer the scenario probability distribution corresponding to the observed outage under this decision. This updates the Dirichlet prior via Bayes’ rule, yielding a posterior mean $\overline{\boldsymbol{p}}_t$, which is then used to define the ambiguity set for the next iteration.}

\textcolor{black}{By employing Bayesian Inference, the posterior distribution from one decision is used as the global prior in the subsequent iteration to represent an updated belief about the uncertainty. A critical aspect of the  online learning algorithm is its ability to address decision-dependent uncertainty by constructing data-driven and decision-specific ambiguity sets $\mathcal{P}_{t+1}(\boldsymbol{h}_t)$, while iteratively learning decisions based on incoming data. The data learned under one decision not only updates the
posterior and ambiguity set for this decision, but also influences other  
decisions through the global prior, facilitating the knowledge transfer between decisions. Furthermore, as more data is incorporated, the empirical distribution becomes increasingly precise, and the distance of the ambiguity set shrinks. This reduced distance signifies that the ambiguity set aligns more closely with the true underlying distribution, mitigating over-conservatism and enhancing decision-making accuracy. This shrinking distance can be modeled as $d_t = \sqrt{{2 |\mathcal{S}| \log\left(\frac{2}{\delta_t}\right)}/{t}}
$, where $\delta_t = \frac{6\delta}{\pi^2 t^2}
$ \cite{rahimian2019distributionally}. Since the algorithm processes outage records sequentially, one data instance per iteration, the iteration index $t$ also represents the cumulative number of observed samples. Note that $\delta_t$
ensures the confidence level, initially $1-\delta$, is maintained over time. Also, the distance selection is based on the $\ell_2$-norm, which guarantees that the true scenario distribution lies within $\mathcal{P}_t$ with probability at least $1 - \delta_t$. An important insight of the online learning algorithm is that a hardening decision impacts only the uncertainty of damage scenarios, rather than the outcomes themselves. As a result, the ambiguity sets for all decisions are equivalent at each iteration. This simplifies the algorithm by requiring only a single ambiguity set to be maintained during the iterative process, significantly reducing computational complexity.}

\begin{algorithm}[t]\small
\caption{\small Bayesian Online Learning Algorithm}\label{A1}
\begin{algorithmic}[1]
\Require Cost function $f(\boldsymbol{h}, s)$, budget $B^{\text{max}}$, and real-world data.
\Ensure Hardening decision $\boldsymbol{h}_T$.
\State \textbf{Initialization:}
\State Set $t = 0$ and initialize hardening decision $\boldsymbol{h}_0$ arbitrarily.
\State Initialize Dirichlet prior: 
\[
\boldsymbol{p}_0 = \text{Dir}(\mathcal{F}_1, \mathcal{F}_2, \ldots, \mathcal{F}_{|\mathcal{S}|}),
\]
where $\mathcal{F}_i=1$ for a uniform prior.
\State Construct ambiguity set $\mathcal{P}_0$ centered at $\boldsymbol{p}_0$ with radius $d_0$.
\For{$t = 1, 2, \ldots, T$}
    \State \textbf{Inner Maximization:} Identify the worst-case  distribution
    \State \quad Compute the gradient ascent update:
    \[
    \tilde{\boldsymbol{p}}_t = \boldsymbol{p}_{t-1} + \eta \nabla_{\boldsymbol{p}}\, \mathbb{E}_{s \sim \boldsymbol{p}_{t-1}} \left[ f(\boldsymbol{h}_{t-1}, s) \right].
    \]
    \State \quad Project $\tilde{\boldsymbol{p}}_t$ onto $\mathcal{P}_{t-1}$ to ensure feasibility:
    \[
    \boldsymbol{p}_t = \arg \min_{\boldsymbol{p} \in \mathcal{P}_{t-1}} \frac{1}{2} \| \boldsymbol{p} - \tilde{\boldsymbol{p}}_t \|_2^2.
    \]
    \State \textbf{Outer Minimization:} Update  hardening decision 
    \[
    \boldsymbol{h}_t = \arg \min_{\boldsymbol{h} \in \mathcal{H}} \mathbb{E}_{s \sim \boldsymbol{p}_t} \left[ f(\boldsymbol{h}, s) \right].
    \]
    \State \textbf{Scenario Translation:} Transform outage scenarios
    \State 
    \[
    \boldsymbol{o}' \leftarrow f_w (\boldsymbol{h}_t, \boldsymbol{c}, \boldsymbol{o}).
    \]
    \State \textbf{Ambiguity Set Update:} Update the probability distribution.
    \State \quad Update the Dirichlet parameters with fractional counts:
    \[
    \boldsymbol{p}_t(D_t, \boldsymbol{h}_t) = \text{Dir}\Big(\mathcal{F}_1 + o'_1,\ \ldots,\ \mathcal{F}_{|\mathcal{S}|} + o'_{|\mathcal{S}|}\Big).
    \]
    \State \quad Compute the posterior mean for each scenario:
    \[
    \overline{\boldsymbol{p}}_t = \left[\frac{\mathcal{F}_1 + o'_1}{\sum_{j=1}^{|\mathcal{S}|} (\mathcal{F}_j + o'_j)},\, \ldots,\, \frac{\mathcal{F}_{|\mathcal{S}|} + o'_{|\mathcal{S}|}}{\sum_{j=1}^{|\mathcal{S}|} (\mathcal{F}_j + o'_j)} \right].
    \]
    \State \quad Update the ambiguity set for the next iteration as:
    \[
    \mathcal{P}_{t+1}(\boldsymbol{h}_t) = \Big\{ \boldsymbol{p} \in \mathbb{R}^{|\mathcal{S}|} : \| \boldsymbol{p} - \overline{\boldsymbol{p}}_t\|_{2} \leq d_{t+1} \Big\},
    \]
    with the confidence radius $d_t = \sqrt{{2 |\mathcal{S}| \log\left({2}/{\delta_t}\right)}/{t}}$
\EndFor
\State \textbf{Output:} Return the final hardening decision $\boldsymbol{h}_T$.
\end{algorithmic}
\end{algorithm}

\subsection{Convergence Analysis Using Dynamic Regret}

To demonstrate the performance of the proposed Bayesian online learning algorithm, its convergence is analyzed. Specifically, the dynamic regret is applied to evaluate the cumulative performance gap between the worst-case performance and the optimal worst-case performance of solving the DRO at each time step $t$. The dynamic regret can be stated as
\begin{equation}\label{DR}
    \mathcal{D}_T = \frac{1}{T} \sum_{t=1}^T \Big( \mathbb{E}_{s \sim \boldsymbol{p}_t} [f(\boldsymbol{h}_t, s)] - \mathbb{E}_{s \sim \boldsymbol{p}_t} [f(\boldsymbol{h}^*_t, s)]+\epsilon_t \Big),
\end{equation}
where $\boldsymbol{p}_t$ is the worst-case distribution. And, $\boldsymbol{h}_t$ represent the hardening decision at iteration $t$ produced by the algorithm, $\boldsymbol{h}^*_t$ represent the optimal decision for the same ambiguity set $\mathcal{P}_t$. As the online algorithm include the DNN regression model in the loop, the prediction error can propagate through the empirical distribution $\hat{\boldsymbol{p}}_{t}$. To evaluate its impact on the convergence, a residual $\epsilon_t$ which quantifies the prediction error is added in equation (\ref{DR}). Note that $\epsilon_t = \|\hat{\boldsymbol{p}}_{t} - \boldsymbol{p}_t^*\|_2$, where $\boldsymbol{p}_{t}^*$ is the true distribution. Based on the fact that the residual $\epsilon_t$ is bounded by the shrinking distance $d_t$ of the ambiguity set, i.e., $d_t = O\left(\sqrt{{\log(t)/}{t}}\right)
$, and the empirical distribution stabilizes as more data is learned over time, hence the cumulative effect of residuals across iterations asymptotically decays. 

\textit{Theorem}: The dynamic regret is bounded:
\begin{equation}\label{DRB}
    \mathcal{D}_T \leq B \left( \sqrt{L/T} + O\left(\sqrt{\log(T)/T}\right) \right),
\end{equation}
where $B$ is the bound for function $f(\boldsymbol{h}, s)$, i.e., $f(\boldsymbol{h}, s)\leq B$. $L$ is the cumulative path length of the ambiguity sets $\mathcal{P}_t$ over all iterations, defined as $L = \sum_{t=1}^T \|\boldsymbol{p}_t - \boldsymbol{p}_{t-1}\|_2^2$. Equation (\ref{DRB}) implies that the regret decreases sublinearly with $T$, ensuring convergence as the algorithm progresses. In other words, the hardening decision derived by the proposed online algorithm become increasingly closer to the exact DRO solutions as more data is learned sequentially.

\textit{Proof}: Using the triangle inequality and the boundedness of function $f(\boldsymbol{h}, s)$, the deviation at each step is stated as:
\begin{equation}
\mathbb{E}_{s \sim \boldsymbol{p}_t} [f(\boldsymbol{h}_t, s)] - \mathbb{E}_{s \sim \boldsymbol{p}_t} [f(\boldsymbol{h}^*_t, s)]+\epsilon_t \leq B \|\boldsymbol{p}_t - \boldsymbol{p}^*_t\|_2,
\end{equation}
where $\boldsymbol{p}_t$ and $\boldsymbol{p}^*_t$ denote the worst-case distribution by using the online algorithm and solving the DRO exactly at time step $t$, respectively. Then, summing over all iterations, the dynamic regret can be derived as
\begin{equation}
    \mathcal{D}_T \leq \frac{1}{T} \sum_{t=1}^T B \|\boldsymbol{p}_t - \boldsymbol{p}^*_t\|_2 + \frac{1}{T} \sum_{t=1}^T \epsilon_t.
\end{equation}

Since the term \( \|\boldsymbol{p}_t - \boldsymbol{p}_t^*\|_2 \) can be bounded as 
$
\|\boldsymbol{p}_t - \boldsymbol{p}_t^*\|_2 \leq \|\boldsymbol{p}_t - \boldsymbol{p}_{t-1}\|_2 + \|\boldsymbol{p}_{t-1} - \boldsymbol{p}_t^*\|_2
$, the right hand side becomes:
\begin{equation}
   \frac{1}{T}B(\sum_{t=1}^T \|\boldsymbol{p}_t - \boldsymbol{p}_{t-1}\|_2 + \sum_{t=1}^T \|\boldsymbol{p}_{t-1} - \boldsymbol{p}_t^*\|_2)+\frac{1}{T} \sum_{t=1}^T \epsilon_t.
\end{equation}

For the first term in the bracket, using the Cauchy-Schwarz inequality, the following inequality can be derived,
\begin{equation}
    \sum_{t=1}^T \|\boldsymbol{p}_t - \boldsymbol{p}_{t-1}\|_2 \leq \sqrt{T \sum_{t=1}^T \|\boldsymbol{p}_t - \boldsymbol{p}_{t-1}\|_2^2},
\end{equation}
which can simply rewritten as $
    \sum_{t=1}^T \|\boldsymbol{p}_t - \boldsymbol{p}_{t-1}\|_2 \leq \sqrt{T L}$.

For the second term in the bracket, since  $\|\boldsymbol{p}_{t-1} - \boldsymbol{p}_t^*\|_2 \leq d_t$, where
$
d_t = \sqrt{\frac{2 |\mathcal{S}| \log\left(\frac{2}{\delta_t}\right)}{t}},  \delta_t = \frac{6\delta}{\pi^2 t^2}$, we have
\begin{equation}
    \sum_{t=1}^T d_t \leq  4\sqrt{2 |\mathcal{S}| T} \left( \sqrt{\log(T)} + \sqrt{\log\left(\frac{\pi^2}{3\delta}\right)} \right).
\end{equation}

The term $\frac{1}{T} \sum_{t=1}^T \epsilon_t$ is the average residual. As  $\epsilon_t = \|\hat{\boldsymbol{p}}_{t} - \boldsymbol{p}_t^*\|_2$ is bounded by the distance $d_t$, by substituting $d_t$ and $\delta_t$, the bound for the avarage residual over time can be derived:

\begin{equation}
    \frac{1}{T} \sum_{t=1}^T \epsilon_t \leq \frac{\sqrt{2 |\mathcal{S}|} \cdot (\log(T))^{3/2}}{1.5 T},
\end{equation}
which means the asymptotic bound on the average residual is $O\left({(\log(T))^{3/2}}/{T}\right)$, which is faster than $O\left({1}/{\sqrt{T}}\right)$.

Combining these results, the final dynamic regret is
\begin{equation}
    \mathcal{D}_T \leq B \left( \sqrt{L/T} + O\left(\sqrt{\log(T)/T}\right) \right),
\end{equation}
where we omit the Big-$O$ term for the average residual, since its contribution is asymptotically dominated by others.

\begin{figure*}[t]
    \centering	\includegraphics[width=6.8in]{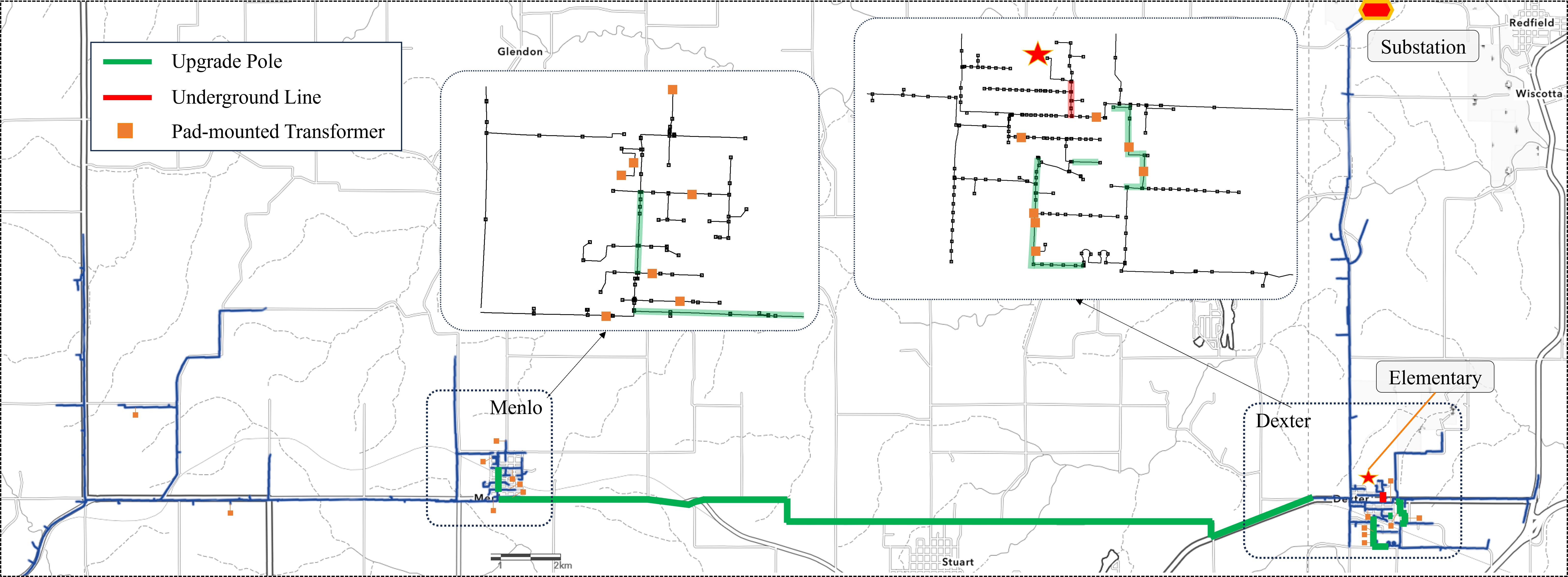}	
    \caption{A geographic illustration of the Redfield distribution network.}
    \label{Redfield}\vspace{-5pt}
\end{figure*}

\subsection{Scalability Analysis}

\textcolor{black}{The proposed strategy is mathematically scalable and practically applicable to much larger systems in both network size and scenario complexity. The computational cost each iteration is dominated by four components: DNN inference, ambiguity set update, projected gradient ascent for the inner problem, and the outer mixed-integer second-order cone programming. The first three are lightweight and scale linearly with the number of scenarios $\mathcal{O}(|\mathcal{S}|)$. Specifically, the DNN inference produces a scenario probability vector; the Bayesian update involves simple Dirichlet count increments; and projection onto the $\ell_2$-norm ambiguity set is a convex optimization problem. Thus, the projected gradient update similarly incurs low cost due to convexity. The outer problem is the most computationally intensive step, yet it is solved by using  commercial solvers such as Gurobi, which employs convex relaxations and presolving to enhance efficiency. We want to note that the entire learning and optimization process is performed offline as part of long-term infrastructure planning. Unlike real-time restoration tasks, hardening plans are determined in advance. Therefore, even for systems comprising thousands of nodes and hundreds of candidate hardening options, the strategy remains tractable.}

\vspace{-5pt}
\section{Case Study}

In this section, the test system is set up. The DNN regression model is trained using real-world outage data. The simulation results of Bayesian online learning alogorithm and data-driven stochastic DL hardening strategy are presented. The comparisons between different hardening strategies are discussed.

\vspace{-5pt}
\subsection{Test System and Data Preparation}

\subsubsection{Test System Setup}

In this paper, a real-world $24.9$kV power distribution network located in Redfield, Iowa, USA is selected for simulation and case study. The  coordination is $<41^\circ 35' 36.1'' \text{N},\ 94^\circ 13' 37.3'' \text{W}>$. The network features a single feeder with a U-shaped configuration. It serves two major load groups located in Dexter, Iowa, and Melon, Iowa, repspectively. An illustration of the distribution network is depicted in Fig. \ref{Redfield}. Specifically, the total length of the main feeder is $47.54$ miles, including 597 nodes, 226 distribution transformers, 19 switches, 1 circuit breaker, 2 recloser, and 60 fuses. The total load demand are $1,813$kW and $1,683$kVar, which are obtained from the utility based on historical electricity consumption. The outage data, spanning from 2001 to 2024, is collected by the utility via the SCADA systems. This dataset contains time stamps, outage duration, and clearing devices. Examples of outage scenarios are listed in Table \ref{Outage_data}. Weather conditions including wind speed and direction, humidity, and temperature, are obstained from the Iowa Environmental Mesonet \cite{Mesonet}. The hardening cost for main feeders is \$0.3M/mile for upgrading poles \cite{Polecost}, and \$3.0M/mile for undergrounding lines \cite{Undercost}, respectively. The hardening cost for laterals is \$0.05M for pad-mounted transformers \cite{PGECOST}. In addition, an Intel Xeon W-1370 workstation, with 2.90 GHz and 32 GB RAM, and Python 3.13, TensorFlow 2.16 and Gurobi solver 12.0 are used as a test platform.

\begin{table}[!t]
\renewcommand{\arraystretch}{1.3}
\caption{Real-world Outage Data Structure}
\label{Outage_data}
\centering
\begin{tabular}{c c c c c c}
\hline
$s$ & Time & Clear Devices & S (mph) & H (\%) & T (°C) \\
\hline
1 & 2015-05-12 14:32 & Transformer & 30 & 65 & 20.9 \\
2 & 2017-08-25 17:15 & Recloser & 40 & 74 & 27.7 \\
3 & 2019-11-03 09:45 & Fuse & 25 & 62 & 18.2 \\
4 & 2021-03-15 12:05 & Transformer & 35 & 68 & 12.3 \\
5 & 2023-07-07 19:20 & Fuse & 45 & 93 & 30.1 \\
\hline
\end{tabular}
\end{table}

\subsubsection{DNN Training and Validation}

\textcolor{black}{Before training the DNN, the raw outage dataset are preprocessed through data augmentation. Specifically, for weather conditions, subtle perturbations are introduced by adding Gaussian noise with a small standard deviation. Similarly, for outage events, minor variations are introduced by adjusting the parameters of clearing devices to mimic those of adjacent devices, thereby accounting for spatial uncertainties.} To train the DNN  model, the outage dataset $\mathcal{T}$ is randomly divided into three subsets, including a training set for optimizing the parameters $w$, a validation set   to fine-tune hyperparameters, and a test set to assess the DNN’s performance. The DNN model consists of three hidden layers, with each layer containing 64 neurons. The Rectified Linear Unit (ReLU) is selected as the activation function. The model is trained with a mini-batch size of 32 samples for up to 100 epochs using the Adam optimizer with an initial learning rate of 0.001. Early stopping was applied by monitoring the validation loss%, terminating training if no improvement was observed for 10 consecutive epochs
. Also, the categorical cross-entropy is adopted as the loss function to measure the difference between the DNN regression model’s predicted and labeled distributions of fault scenarios. 

\begin{figure}[t]
	\centering	\includegraphics[width=3.4in]{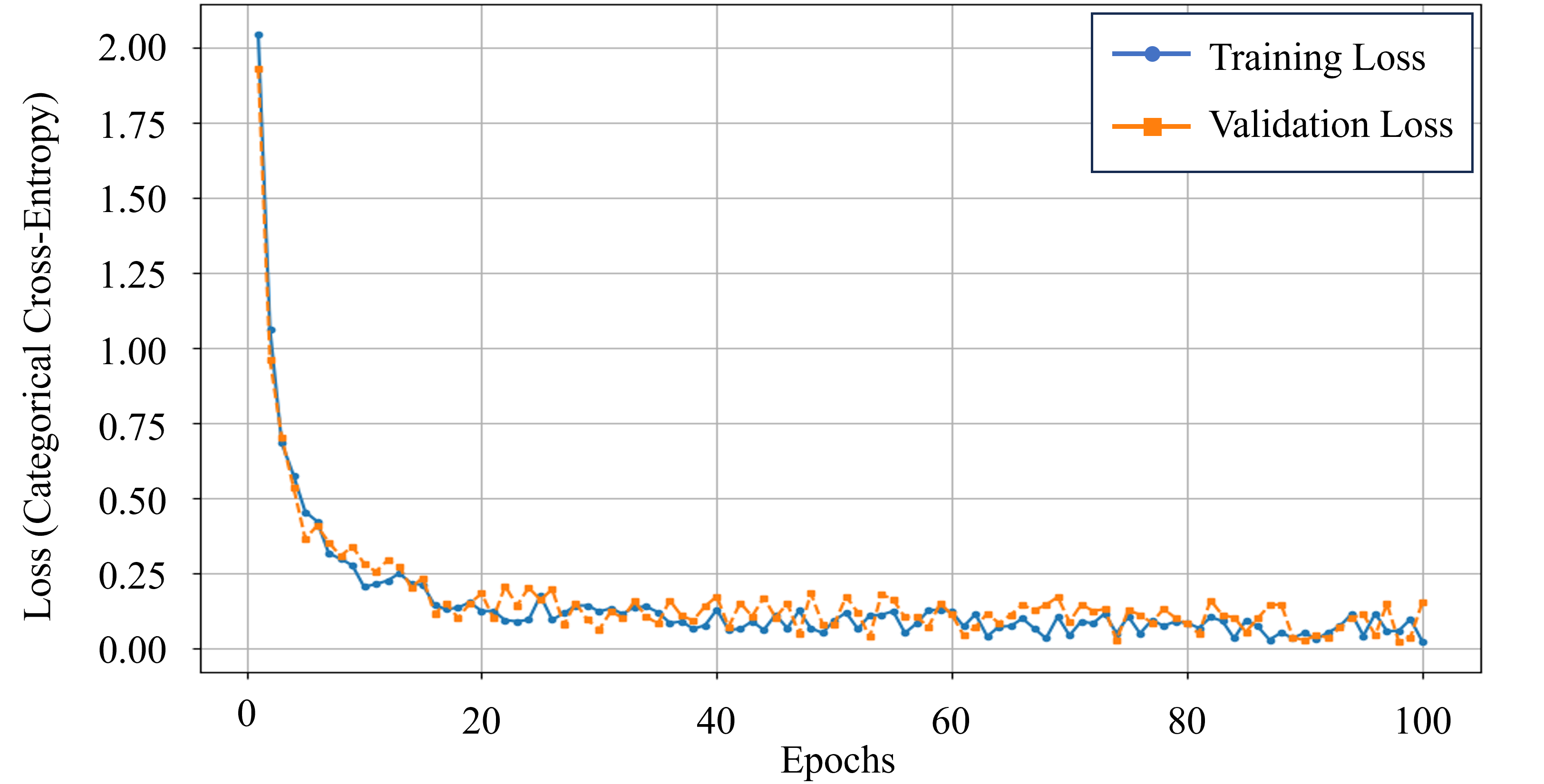}	
	\caption{An illustration of the learning process of the DNN model.}
	\label{DNN}\vspace{-5pt}
\end{figure}

Fig. \ref{DNN} depicts the training loss and validation loss curves. We can see both the loss curves decrease obviously, which effectively demonstrate the  learning process. In particular, the validation loss stabilizes around epoch 45. It means  the DNN model converges with minimal overfitting, and the validation performance does not deteriorate despite additional training. Furthermore, we evaluate the DNN model based on the test set. The results show the model achieves an accuracy of 92\%, with a precision of 90\% and a recall of 88\%, to predict the outage scenario with the highest probability. In addition, the predicted probability distributions over all outage scenarios were assessed, resulting in a mean absolute error of 0.05 and a root mean squared error of 0.07. These results confirm that the proposed DNN model accurately captures the evolution of outage scenarios under different hardening measures.

\vspace{-5pt}

\vspace{-5pt}

\subsection{Bayesian Online Learning Implementation}

\begin{figure}[t]
	\centering	\includegraphics[width=3.4in]{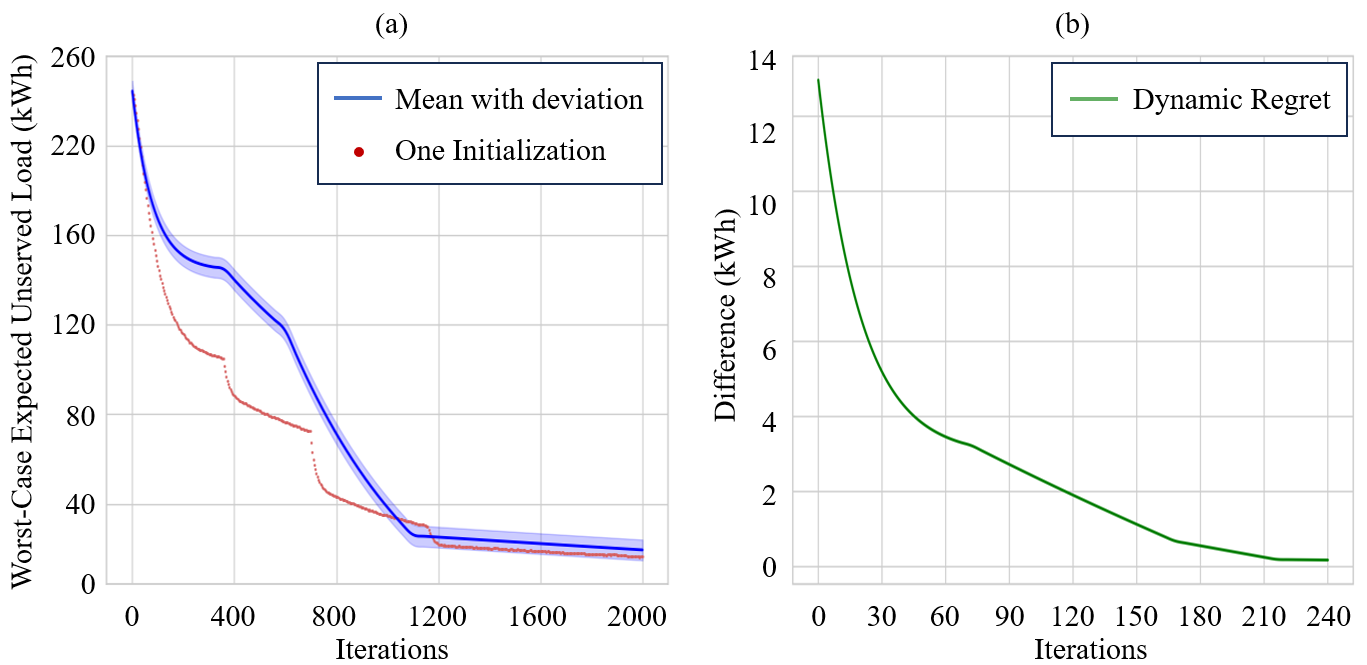}
	%\vspace{-5pt}
	\caption{An illustration of Bayersian online learning process.}
	\label{Online}%\vspace{-5pt}
\end{figure}

In this section, we evaluate the performance of the  Bayesian online learning algorithm. The results are averaged over $50$ random  initializations, with both the mean and standard deviation plotted to illustrate the stability of the algorithm. In Fig. \ref{Online}(a), the blue solid line represents the evolution of the average worst-case expected load shed over 2,000 iterations. Initially, the mean value is very high because the ambiguity set is large due to limited data, meaning the empirical distribution remains largely random. Accordingly, the hardening decisions at the beginning  can be overly conservative. Nevertheless, as iterations progress, more data are incorporated and learned, the empirical distribution becomes more reliable and the ambiguity set naturally shrinks. Consequently, the worst-case expected unserved load exhibits a steady decline. And, after iteration $1100$, the objective value stabilizes around around $20.00$kWh. It implies the hardening decisions become less conservative and more consistent with the actual outage uncertainty. The red dot line in Fig. \ref{Online}(a) shows the online learning process of one initialization. The results reveal three abrupt improvements at iteration $367$, $691$, and $1314$. These reflect the mixed-integer nature of the  proposed hardening problem. Also, the algorithm demonstrates the ability to escape local optima and explore superior solutions. \textcolor{black}{In addition, Fig. \ref{Online}(b) presents the convergence of the dynamic regret, which quantifies the performance gap between the online learning algorithm and the exact DRO solution at each iteration. The fact that the dynamic regret tends to zero after approximately 240 iterations indicates that the algorithm has accumulated sufficient information to generate solutions that are effectively equivalent to those of the exact DRO problem with the current data. In other words, from that point onward, the online learning algorithm is producing decisions that are reliably near-optimal for the evolving problem. However, achieving near-zero dynamic regret does not imply convergence of the objective value. Since the ambiguity set continues to be refined as new data are incorporated, the worst-case expectation may continue to change. This process continues until around iteration 1100, when both the ambiguity set and the decision vector $\boldsymbol{h}_t$ cease to change meaningfully.} To further demonstrate the computational performance, we conduct comparisons between the Bayesian online learning algorithm and the exact DRO in terms of the average computational times per iteration. The comparisons are based on $10$, $25$, $50$, and $100$ randomly selected outage scenarios. The results listed in Table \ref{Computation} indicate that the proposed online learning algorithm requires substantially less computational time per iteration, especially as the number of scenarios increases. This efficiency is primarily due to our decomposition strategy, which bypasses the need to solve the complete DRO problem in each round.

\begin{figure}[t]
	\centering	\includegraphics[width=3.4in]{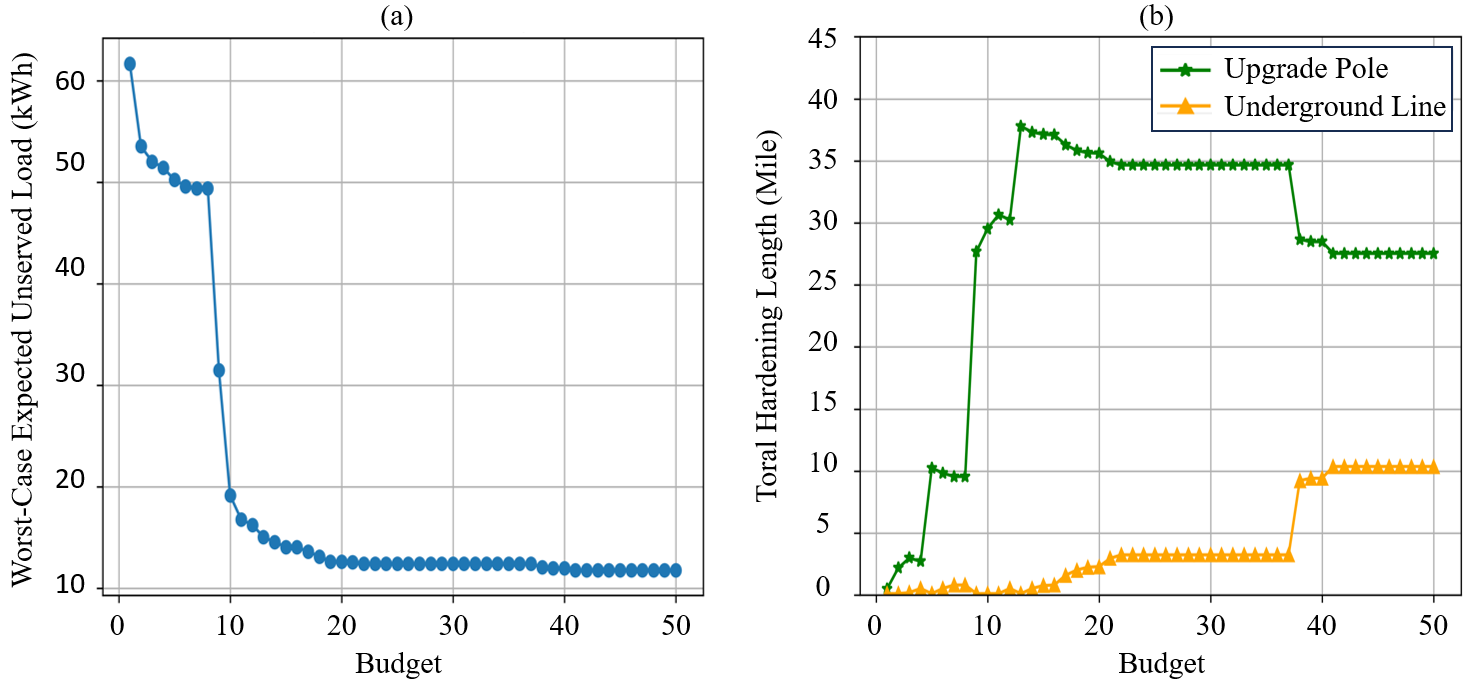}
	%\vspace{-5pt}
	\caption{Results of varying budget in terms of hardening outcomes.}
	\label{Numerical}
\end{figure}

\subsection{Numerical Results and Discussion}

In this section, we discuss about the numerical results for the Redfield distribution network under both fixed and variable budget scenarios. For the fixed case, a total hardening budget of \$10.0M is considered. Then, we determine the hardening decision through the Bayesian online learning algorithm. The results show that a combined solution of $29.5$ miles of pole upgrading, $0.114$ miles of undergrounding, and the reinforcement of $16$ distribution transformers can minimize the worst-case expected unserved load to $19.17$kWh. Fig. \ref{Redfield} illustrates the hardening solution. First, the components whose failures can result in significant losses are hardened. For instance, the long line segment connecting the two concentrated load groups with a length of $9.18$ miles is prioritized for hardening. This result is reasonable, as protecting critical links between major load areas is essential. If this line segment is not hardened as suggested, the worst-case expected unserved load will increase to $92.09$kWh. Also, under the $\$10.0$M fixed budget, this long line segment can only be hardened through upgrading poles due to the budget limits. Second, the solution prioritizes the uninterrupted service of critical loads. For instance, a key line segment supplying power to the West Elementary School is specifically undergrounded, demonstrating the proposed strategy can effective incorporate real-world outage data and allocate resources efficiently. If this line segment is not hardened as suggested, the worst-case expected unserved load would increase to $48.20$kWh. Third, distribution transformers with substantial unserved load scenarios are replaced with pad-mounted transformers. Furthermore, to evaluate the impact of DNN prediction accuracy, we perform a sensitivity analysis by perturbing the DNN output probabilities with zero‑mean Gaussian noise with a standard deviation of 10\%, followed by renormalization. The hardening strategy is then re-executed across 50 such perturbed scenarios. Simulation results show that the average worst-case load shed increases by only 3.59\%, while the selected hardening lines maintains a 93.8\% overlap with the original plan. These findings indicate that the DRO-based ambiguity set offers robustness against moderate prediction errors, preserving decision quality under uncertainty.

For the variable budget case, the comparative results are shown in Fig.  \ref{Numerical}. Specifically, as the budget increases, the worst-case expected unserved load is decreased. It demonstrates that higher investments enhance system resilience. In Fig. \ref{Numerical}(b), at higher budget levels, it is obvious that the solution increasingly prioritizes undergrounding over pole upgrading. This shift occurs because  undergrounding substantially reduces outage probabilities. Conversely, when financial resources are limited, pole upgrading remains a more cost-effective alternative. For example, when the budget exceeds \$15M, sufficient financial resources become available for undergrounding. However, the reduction in the expected unserved load beyond this threshold is less pronounced. This finding highlights the importance of optimal balancing pole upgrading and undergrounding rather than assuming that increased undergrounding always yields the most resilient measures.

\begin{table}[t]
	\renewcommand{\arraystretch}{1.3}
	\caption{Computational Performance of Online Learning}
	\label{Time}
	\centering
	\begin{tabular}{c c c c}
		\hline
		$|\mathcal{S}|$ &  Online Learning & DRO Each Step & Efficiency Improvement  \\
		\hline
		10  & 1.22s  & 2.34s & 47.86\%\\
		25  & 3.61s  & 7.64s & 52.75\% \\
		50  & 12.3s  & 32.0s & 61.56\%\\
		100  & 21.5s  & 85.6s & 74.88\%\\
		\hline
	\end{tabular}\label{Computation}
\end{table}

\vspace{-5pt}

\subsection{Comparative Study}

To futher demonstrate the advantages of the proposed data-driven stochastic DL hardening strategy, we compare its performance against the following strategies:

\begin{enumerate}
    \item  Robust optimization \cite{lin2018tri}: Minimizes the worst-case unserved load considering the most severe outage scenario.

    \item Stochastic programming \cite{ma2016resilience}: Minimizes unserved load based on the expected performance over all scenarios.

    \item DRO \cite{li2023distributionally}: Minimizes the worst-case expected unserved load by considering distributions using an ambiguity set.
\end{enumerate}

\begin{figure}[t]
	\centering	\includegraphics[width=3.3in]{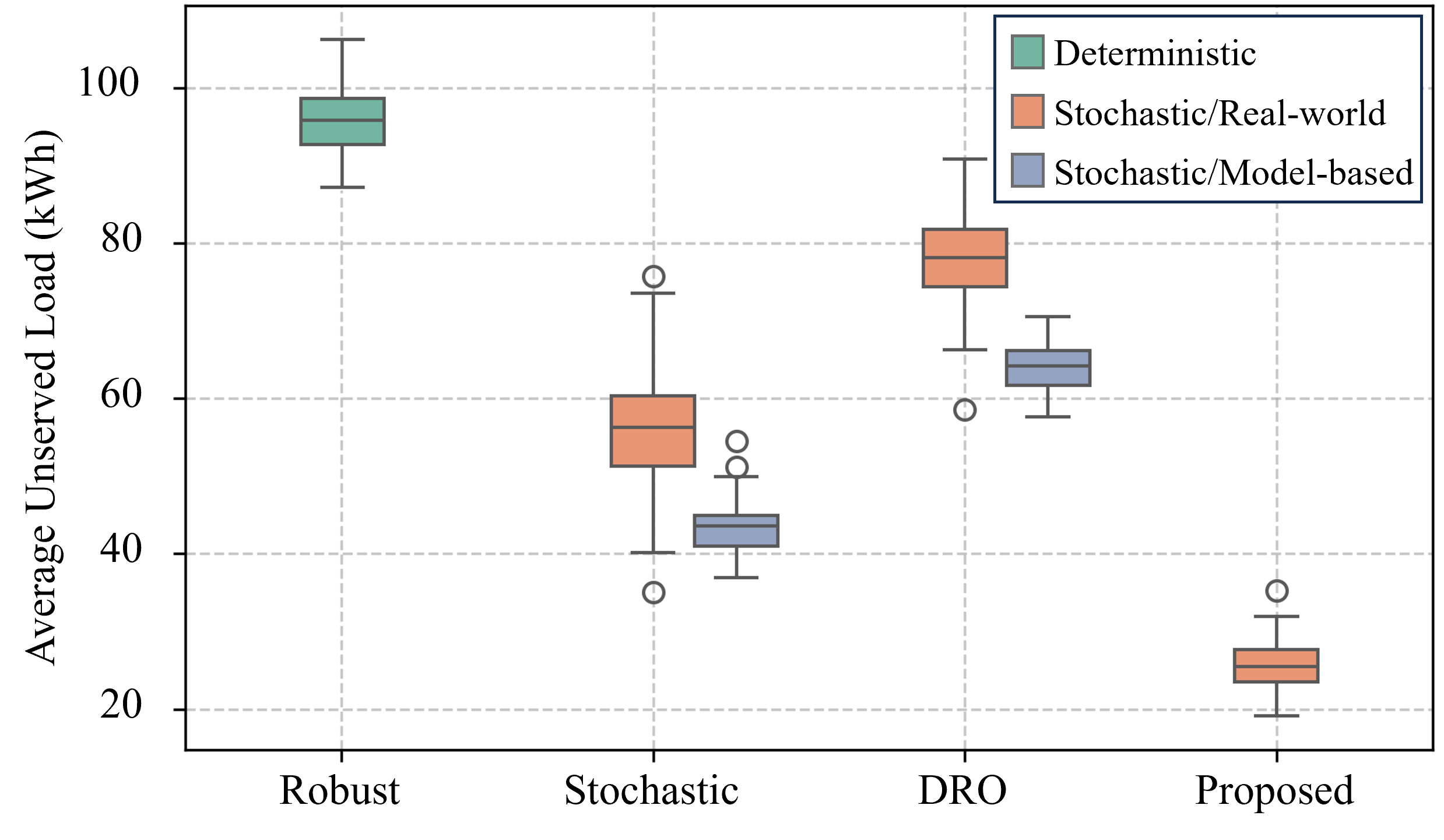}
	%\vspace{-5pt}
	\caption{Comparative results between  hardening strategies.}
	\label{Comparison}\vspace{-5pt}
\end{figure}

Among these strategies, robust optimization is a deterministic approach without consideration of probabilities. In contrast, both stochastic programming and DRO derive the probability distributions of outage scenarios from mathematical models, while our proposed data-driven strategy leverages real-world outages. For the comparison, each hardening strategy is evaluated based on multiple trials. Specifically, for each hardening decision of each strategy, we implement 50 independent trials, each trial includes 50 randomly generated scenarios. For each scenario, the unserved load is calculated in terms of the fixed hardening decision, and the average unserved load for the trial is computed. This procedure is repeated to construct a distribution of average unserved load values. From this distribution, key statistical measures, including the mean, upper bound, and lower bound, are derived, as presented in Fig. \ref{Comparison}.

From these results, we can observe that hardening strategy using robust optimization derives the highest average unserved load, which is $96.31$kWh. The reason is that robust optimization explicitly protects against the worst-case scenario. In contrast, stochastic programming  maximizes expected performance across all scenarios, which results in a lower average unserved load of $57.11$kWh. However, we can see the discrepancy between the mean and upper is bound is larger in terms of real-world scenarios and model-based scenarios. It means that stochastic programming can underestimate risk if the model-based distribution deviates from actual outage behavior. In addition, the results of DRO lies between RO and SP, offering a balance between conservatism and adaptability. This is achieved by considering an ambiguity set of distributions. However, if the ambiguity set is derived from an inaccurate model, discrepancies can still arise when evaluated against real-world data. Furthermore, the proposed data-driven stratey achieves the best overall performance with the lowest average unserved load of $24.05$kWh. The reason is its ability to continuously update the outage distributions based on online learning of real-world outages, capturing the stochastic nature of system failures more accurately.

\section{Conclusion}

In this paper, a data-driven stochastic DL hardening strategy is proposed to leverage real-world data for making informed hardening decisions. By integrating a DNN regression model for outage distribution prediction with a decision-dependent DRO model, the proposed strategy effectively incorporates historical outage information into a data-driven ambiguity set. A Bayesian online learning algorithm is proposed to address the decision-dependent uncertainty and enhance
computational efficiency. This algorithm decompose the problem into inner and outer problems based
on Bayes’ Theorem and Bayesian Inference, then iteratively updates the empirical distribution and the hardening decisions by sequential  learning. Simulation results on a real-world Redfield distribution network demonstrate the strategy’s capability to minimize worst-case expected unserved load while outperforming model-based strategies.

Future research can be extended in several promising directions. First, the proposed strategy may be enhanced by incorporating high-frequency phasor measurement unit (PMU) data to capture transient phenomena such as cold load pickup following extended outages. This integration would enable more accurate modeling of load recovery dynamics and further strengthen distribution system resilience. \textcolor{black}{Second, to address the computational challenges posed by the mixed-integer optimization in the outer loop, learning-based surrogate models or neural warm-start generators can be explored. These tools have the potential to produce high-quality initial solutions, reducing solution times and improving overall scalability.}

\bibliographystyle{IEEEtran}\bibliography{ref}

\end{document}